\documentclass[runningheads]{llncs}
\usepackage{comment}
\usepackage[textsize=tiny]{todonotes}
\usepackage{hyperref}
\usepackage{amsmath,amsfonts}
\usepackage[capitalize,noabbrev]{cleveref}
\usepackage{enumitem}
\usepackage{tabu}
\usepackage{textcomp}
\usepackage[utf8]{inputenc}
\usepackage[T1]{fontenc}
\usepackage{microtype}
\usepackage[a-2b]{pdfx}
\usepackage{graphicx}
\usepackage{balance}
\usepackage{subcaption}
\usepackage{multirow}
\usepackage{listings}
\usepackage{float}
\usepackage{xcolor}
\usepackage{outlines}
\usepackage[normalem]{ulem}
\usepackage{cleveref}
\usepackage{soul}
\usepackage{cuted, lipsum}
\usepackage[listings,skins,breakable]{tcolorbox}
\usepackage{booktabs}
\newcommand{\eat}[1]{}
\begin{document}

\title{Declarative Privacy-Preserving Inference Queries}

\titlerunning{Declarative-Privacy}

\author{Hong Guan, Ansh Tiwari, Summer Gautier, Rajan Hari Ambrish, \\ Lixi Zhou, Yancheng Wang, Deepti Gupta$^\dagger$, \\Yingzhen Yang, Chaowei Xiao$^\ddagger$, Kanchan Chowdhury, Jia Zou}

\institute{Arizona State University, Texas A\&M University-Central Texas$^\dagger$, University of Wisconsin at Madison$^\ddagger$}






\maketitle

\begin{abstract}
Detecting inference queries running over personal attributes and protecting such queries from leaking individual information requires tremendous effort from practitioners. To tackle this problem, we propose an end-to-end workflow for automating privacy-preserving inference queries including the detection of subqueries that involve AI/ML model inferences on sensitive attributes. Our proposed novel declarative privacy-preserving workflow allows users to specify \textit{``what private information to protect''} rather than \textit{``how to protect''}. Under the hood, the system automatically chooses privacy-preserving plans and hyper-parameters. Link to our video: \url{https://youtu.be/nK2deY_6adM}
\end{abstract}

\section{Introduction}
\label{sec:intro}
With more database systems supporting AI/ML, queries increasingly involve AI/ML model inferences. However, limited research has addressed declarative support for differential privacy (DP) in inference queries, particularly when the underlying dataset contains sensitive information. Ideally, with declarative DP support, a data owner will declare sensitive information that needs to be protected by tainting the data attributes and tuples. Database users could then issue arbitrary inference queries without needing to define the specifics of the DP mechanism or model. Instead, the system would automatically identify sensitive subqueries and apply DP safeguards based on the user’s privacy budget.
We use an example to illustrate the idea as follows. 

\noindent
\textbf{Motivating Example.} Online social media posts may disclose personal information such as use patterns, life habits, and social status. Linkage attacks can relate the information in the online post to private information. Consider the following query, ``\textit{SELECT count(*) FROM IMDB\_MOVIE\_REVIEW R WHERE R.date > `06/01/2015' AND R.date < `06/05/2015' AND sentiment\_classifier(R.Review) = Positive }''. Instead of directly sharing private reviews with business analysts or data scientists, the social media company can train a sentiment classification model that takes a review within the query range as input and outputs its sentiment. Then the sentiment predictions will be grouped and aggregated. Differential Privacy-Stochastic Gradient Descent (DP-SGD)~\cite{abadi2016deep} ensures, with high probability, that business analysts cannot reconstruct any private training data from the model. As a result, the system will use the privacy-preserving model to answer the query, which often achieves better privacy-utility trade-offs than alternatives, such as adding perturbations to the aggregation results.
The privacy cost of the query is constrained by the remaining privacy budget of both the dataset and the user.
However, there are several new challenges: (1) How to detect the portion of queries that need to be protected? (2) How to search for a neural network architecture that balances the privacy and accuracy, given the training data and validation data and the privacy budget ($\epsilon$)? For example, we may fine-tune a pre-trained BERT model on the private social media posts using DP-SGD, while we can also train a bi-LSTM model with pretrained Word2Vec embeddings from scratch using DP-SGD. Alternatively, a pre-trained deep learning model can be used to encode each data point into an embedding vector, with noise added to ensure differential privacy. In this case, inference is performed using approximate nearest neighbor search.
Using the IMDB dataset, we found that the privacy-utility trade-offs of the first two approaches outperformed the third approach.
The first approach (i.e., fine-tuned a pre-trained BERT model using DP-SGD) achieved better accuracy than the second approach (i.e., training a bi-LSTM model from scratch using DP-SGD) for small privacy budgets (i.e., $\epsilon < 6$), while the second approach outperformed the first approach for other cases. 

To address these challenges, we developed a declarative system that consists of the following components: (1) Taint Analysis, which automatically identifies the sensitive subqueries; (2) Privacy-Preserving Query Transformer, which transforms a user-requested inference operator into an equivalent operator with desired privacy guarantees; (3) Differentiable Neural Architecture Search, which automatically searches for the optimal neural network architecture for training the model to be used by the transformed inference operator, if there are no existing suitable models available. In the rest of the paper, we will introduce each component in Sec.~\ref{sec:architecture} and describe the demonstration proposal in Sec.~\ref{sec:demo-proposal}.

\section{System Architecture}
\label{sec:architecture}
Our proposed system consists of the following components: 


\noindent
\textbf{Query Taint Analysis}
We enhance the data catalog to enable the tainting of private attributes based on users' privacy requirements~\cite{zhou2023privacy}. Attributes to be tainted are directly specified by the data owners or extracted from the view-based access control policy specified by the data owners. Then, once a data scientist issues a query, the query will be lowered to a graph-based Intermediate Representation (IR) based on the nested relational algebra, where each node represents a relational operator, and each edge represents a dataset, which could be a relation/view or a collection of objects, such as images, video, and text files. The tainted sources will propagate the taint through the IR graph, so that the sensitive sub-queries that access private information are identified.

\vspace{3pt}
\noindent
\textbf{Privacy-Preserving Transformation}
We abstract the process of providing DP support for a sensitive query as a special type of query transformation rules, described as follows (using the classical relational algebra notations):
%
(1) $\lambda_{f} R \overset{\Delta{acc}, \epsilon}{=} \lambda_{f'} R$. $\lambda_{f}$ represents the prediction operator that takes each tuple in the relation (or a collection of arbitrary objects) $R$ as input features, and outputs prediction results. The DP-SGD-trained model, denoted as $\lambda_{f'} R$,  outputs the prediction results with privacy guarantee $\epsilon$. In addition, such transformation may results in an accuracy drop, represented as $\Delta acc$.
(2)  $\lambda_{f}(\pi_{A}(\sigma_{p}(R))) \overset{R, \Delta{acc}, \epsilon}{=} \lambda_{f'}(A, p)$. $\lambda_{f}$ represents the prediction operator taking the query output $\pi_{A}(\sigma_{p}(R))$ as input features. The model trained with DP-SGD, denoted as $\lambda_{f'}(A, p)$, which takes the projection attributes $A$ and selection predicate $p$ as inputs, and outputs the final prediction result for $\lambda_{f}(\pi_{A}(\sigma_{p}(R)))$. We have similar rules extended for aggregation queries, which are omitted due to space limitation. A query optimizer searches for the most promising transformation plan using a learned cost model.

\vspace{5pt}
\noindent
\textbf{Differentiable Neural Architecture Search}
To find the optimal architecture of the neural network model with minimal human effort, we adopted a Differentiable Neural Architecture Search (DNAS) algorithm \cite{fbnetv2}. When a most relevant foundation model is found, DNAS will be used to search for layers to be fine-tuned and/or the architecture of new adaptive layers for fine-tuning; otherwise, DNAS will search for the architecture of a new model based on the modality of the data. In DNAS, the choices of various neural operations are characterized by a set of architecture parameters defined in a graph consisting of all possible layers. 
DNAS searches for the optimal network architecture and model weights through gradient descent by optimizing both the prediction loss and the computation cost of the neural networks. 
One challenge is that the model search process will also incur privacy costs. To address the issue, once the search is complete, we \textit{reset the weights of the searched neural network to the initial random state} and train it on the private data using DP-SGD. Note that although DNAS is conducted on private data, it does not increase the privacy budget as the network weights of the supergraph will not be released~\cite{cheng2022dpnas}.

\begin{figure*} 
\centering
\includegraphics[width=1\textwidth]{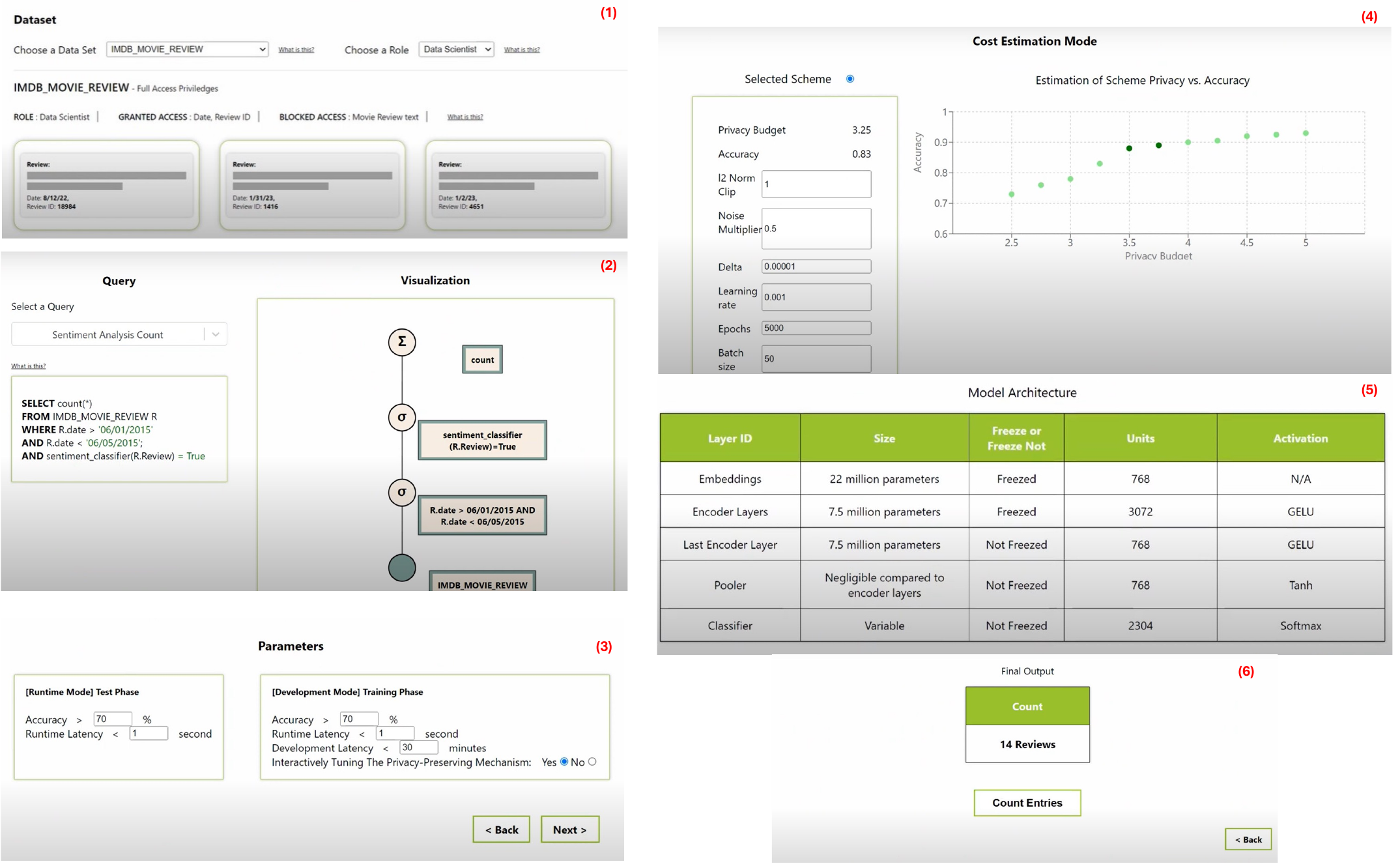}
\caption{\label{fig:UI} \small
Illustration of the Demo and User Interfaces
}
\end{figure*}

\vspace{-5pt}
\section{Demonstration Proposal}
\label{sec:demo-proposal}

We will demonstrate the proposed declarative privacy-preserving workflow. The demo audience can select a database, and then run the following steps:

\noindent
\textbf{Step 1. Data Owners Annotate Sensitive Information.} 
Data owners label the attributes and tuples containing private information and specify the corresponding privacy parameters for different user roles. Such information can also be automatically extracted from the traditional RDBMS view-based access control policies that are defined by the data owners. As illustrated in Figure ~\ref{fig:UI} (1), From the perspective of the data scientist, sensitive information is redacted while only insensitive information is listed. 

\noindent
\textbf{Step 2. Query Request and Analysis.} As illustrated in Figure ~\ref{fig:UI} (2), after the data scientist issues queries over private datasets, an intermediate representation of the query is visualized at the right side of the panel, and the sensitive sub-queries that involve private attributes will be highlighted in green boxes. 

\noindent
\textbf{Step 3. Parameter Specification.}
The system recommends training and test parameters to the data scientist, such as accuracy and latency. The system administrator can change these default settings and choose to interactively tune the parameters, as illustrated in Figure ~\ref{fig:UI} (3).

\noindent
\textbf{Step 4. Privacy-Preserving Recommendation.} As illustrated in Figure ~\ref{fig:UI} (4), the system will visualize the top $k$ privacy-preserving schemes that are recommended by the system. These recommended schemes are sampled from the Pareto-optimal plans.  For each privacy-preserving scheme, it will list the path to the pre-trained model, or the model architecture and hyper-parameters for training a new model, as illustrated in Figure~\ref{fig:UI} (5). The system administrator or a system program can select a configuration according to the task requirement.

\noindent
\textbf{Step 5. Query Execution.} 
The system proceeds to run the query with visualized query results as shown in Figure ~\ref{fig:UI} (6).

The above steps outlined our proposed declarative privacy-preserving workflow for inference query execution in an interactive style. 

\vspace{-10pt}

\bibliographystyle{splncs04}
\bibliography{refs}

\end{document}